# Radiation tests of the Silicon Drift Detectors for LOFT

E. Del Monte[a,b,*], P. Azzarello[c], E. Bozzo[c], S. Bugiel[d], S. Diebold[e],
Y. Evangelista[a,b], E. Kendziorra[e], F. Muleri[a,b], E. Perinati[e], A. Rachevski[f],
G. Zampa[f], N. Zampa[f], M. Feroci[a,b], M. Pohl[g], A. Santangelo[e], A. Vacchi[f]

[a]INAF - IAPS, Via Fosso del Cavaliere 100, I-00133 Roma, Italy
[b]INFN Sezione di Roma Tor Vergata, Via della Ricerca Scientifica 1, I-00133 Roma Italy
[c]ISDC, Université de Genève, Chemin d'Ecogia 16, CH-1290 Versoix, Switzerland
[d]MPIK - Max Planck Institut für Kernphysik, 69117 Heidelberg, Germany
[e]IAAT, Universität Tübingen, Sand 1, 72076 Tübingen, Germany
[f]INFN Sezione di Trieste, Padriciano 99, I-34149 Trieste, Italy
[g]DPNC, Université de Genève, Quai Ernest-Ansermet 24, CH-1211 Genève, Switzerland

**ABSTRACT**

During the three years long assessment phase of the LOFT mission, candidate to the M3 launch opportunity of the ESA Cosmic Vision programme, we estimated and measured the radiation damage of the silicon drift detectors (SDDs) of the satellite instrumentation. In particular, we irradiated the detectors with protons (of 0.8 and 11 MeV energy) to study the increment of leakage current and the variation of the charge collection efficiency produced by the displacement damage, and we "bombarded" the detectors with hypervelocity dust grains to measure the effect of the debris impacts. In this paper we describe the measurements and discuss the results in the context of the LOFT mission.

## 1. INTRODUCTION

LOFT[1] (Large Observatory For x-ray Timing) is a satellite-borne mission devoted to the spectral-timing studies of Astrophysical sources in the X-ray band. LOFT was candidate to the M3 launch opportunity of the European Space Agency under the long-term science plan "Cosmic Vision 2015 – 2025" and underwent a three year long assessment phase[2]. The LOFT instrumentation is composed of the Large Area Detector (LAD[3]) and the Wide Field Monitor (WFM[4]). The LAD is a collimated instrument for X-ray timing with the unprecedented effective area of ~10 $m^2$ at 8 keV energy. The WFM is a coded aperture instrument and is in charge of monitoring the X-ray sky to localise transients, outbursts and other interesting states of known and newly discovered sources. These can be observed with the LAD after repointing the satellite.

Both the LAD and WFM make use of innovative large-area Silicon Drift Detectors (SDDs[5,6,7]). With these sensors we can cover the large geometric area of the LAD (~18 $m^2$) with a relatively small number of tiles achieving a spectral resolution of 200 – 250 eV Full Width at Half Maximum (FWHM) at 6 keV, and a collection time of ~5 μs. During the assessment phase of LOFT, a dedicated team optimised the SDD design[8] in order to improve the performance and fulfil the mission scientific requirements.

*ettore.delmonte@iaps.inaf.it; phone +39 0649934675; fax +39 0645488188

Silicon detectors are particularly sensitive to the displacement damage produced by energetic charged and neutral particles in orbit, affecting the lattice of the semiconductor and increasing the detector leakage current. Such current increase worsens the spectral resolution of the SDDs, thus affecting some of the scientific performances of the instruments. It is particularly relevant for the observation with the LAD of relativistically broadened and skewed profiles of spectral lines (e.g. iron $K_\alpha$) from X-ray binary systems and active galactic nuclei[2] in order to derive the black hole spin. For these observations a spectral resolution of FWHM $\leq$ 240 eV on average is requested. This translates into a requirement on the Equivalent Noise Charge (ENC) of the electronic noise for the LAD[2], $\sigma_{ENC} \leq 17$ e$^-$ root mean square (RMS). Since the SDD leakage current is an important contribution to the electronic noise, the spectral resolution is expected to worsen due to the displacement damage, if not mitigated.

In addition, in order to reach the required sensitivity of the WFM at low energy (e.g. between ~2 keV and ~6 keV) the SDDs should observe the sky with the minimum possible amount of shielding material in the field of view. For this reason the detectors might end up being particularly exposed to the impact of micrometeoroids and orbital debris (MMODs) in orbit. these are hypervelocity particles which are able to damage the sensors upon impact.

During the LOFT assessment phase we measured the effect on the SDDs of the radiation damage from protons and MMODs. In this paper we report about the measurements and summarise the results. The paper is organised as follows: in Sect. 2 we outline the environment of the LOFT mission; in Sect. 3 we deal with the measurements of the radiation damage and in Sect. 4 we report the effect of the impact of hypervelocity particles; in Sect. 5 we summarise the results, explain the mitigation strategy for the LOFT instrumentation and draw our conclusions.

## 2. THE LOFT ENVIRONMENT

### 2.1 Environment of charged particles

By using methods reported in the literature[9] we estimated the increase in leakage current produced by the displacement damage following the Non Ionising Energy Losses (NIEL) scaling hypothesis[10]. In this approximation the damage depends only on the particle fluence and the variation of the damage with the particle type and energy is introduced via a parameter, the hardness factor[10,11]. The displacement damage is generally reduced by the annealing process[12], i.e. the redistribution of the damage centers in the lattice, depending on the time and storage temperature after the irradiation. The additional leakage current produced by the displacement damage shows the same variation as a function of temperature[13,14,15] as the intrinsic, preexisting leakage current of the devices.

During the LOFT study, we selected the altitude and inclination of the satellite orbit in order to minimise the displacement damage. For this purpose we estimated, using the SPENVIS[16] web-based software, the particle environment in Equatorial Low-Earth Orbits (LEOs) with altitude $\leq$ 600 km and inclination $\leq$ 5°. In the estimation we took into account only the displacement damage from protons, because the damage of electrons, about three orders of magnitude smaller than protons, and albedo neutrons, whose fluence rapidly decreases above ~10 MeV, gives a small contribution. We estimated the proton fluence for the LOFT orbits using the AP8[17] model in the worst case condition of Solar Minimum activity (AP8-MIN). SPENVIS allows the user to transport the protons through an aluminum shielding, and to estimate the equivalent fluence at a single energy impinging on the detector. In our estimation we selected a proton energy of 10 MeV.

The lowest fluence is for 550 km altitude and 0° inclination, assumed as the baseline orbit in the LOFT industrial study[2]. Since the flux of AP8-MIN is formally zero at this orbit, we conservatively considered the flux at an inclination of 3.5°. In addition, we applied a margin of 20 times, resulting from the comparison between the flux estimated by AP8-MIN at 550 km and 2° and the flux measured by the Standard Radiation Environment Monitor[18] aboard the Proba-1 satellite at 550 km and 0°. The equivalent fluence of 10 MeV protons at the LOFT baseline orbit is thus $3 \times 10^6$ cm$^{-2}$ (including margin). For comparison, the 10 MeV proton fluence at 600 km and 5° is $1.8 \times 10^9$ cm$^{-2}$ (including margin).

An additional low-energy proton component[19], concentrated at Equatorial latitudes, almost independent on altitude between 500 km and 1000 km and not included in AP8, has been considered in our estimation[20]. This component is characterised by a steep power-law spectrum[19], extending up to energies of several MeV, and a narrow angular distribution[14] proportional to $\sin^6(\theta)$ around the pitch angle $\theta$. Since this soft component is not included in SPENVIS, we calculated the expected increase in leakage current for the particular case of the SDDs in the LAD and WFM[20].

In this energy range the losses are comparable to the proton kinetic energy, and thus some particles are stopped in the material layers preceding the detectors. Since the energy loss is not a linear function of the kinetic energy, evaluating an average amount of equivalent shielding material, as done in SPENVIS, would be rather cumbersome, and certainly it would not be justifiable. Instead, we used the SRIM/TRIM[31] simulation code to propagate the incident soft proton spectrum up to the SDD surface, and then we used the same software to calculate the effective hardness factor at different energies. By using the NIEL scaling hypothesis, we could then estimate the expected total mission soft proton fluence at a reference energy, chosen to be 0.8 MeV. For LOFT the damage is expected to be larger on the WFM detector, since it features a lower level of shielding due to the wider acceptance angle of the coded mask (~2.5 sr) as compared to the LAD collimator (~3 × $10^{-4}$ sr). For a five years mission the resulting conservative estimation of the equivalent fluence, which we used as a baseline to plan the irradiation test, is $3.6 \times 10^7$ cm$^{-2}$.

## 2.2 Environment of micrometeoroids and orbital debris

The flux of micrometeoroids and orbital debris, i.e. the average number of particles per unit time and unit area impinging on a randomly tumbling surface (corresponding to an acceptance angle of $2\pi$ sr), roughly follows a power law distribution[21]. The most abundant particles have typical size between ~1 μm and ~10 μm (see Figure 1). Since the LOFT spacecraft will not stay at a fixed position but will be pointed and oriented according to a pre-defined observation plan, we assume in our analysis an isotropic distribution of MMODs. We considered an average density and velocity of 2.5 g/cm$^3$ and 20 km/s for the micrometeoroids, and 2.8 g/cm$^3$ and 13 km/s for the debris, respectively[21].

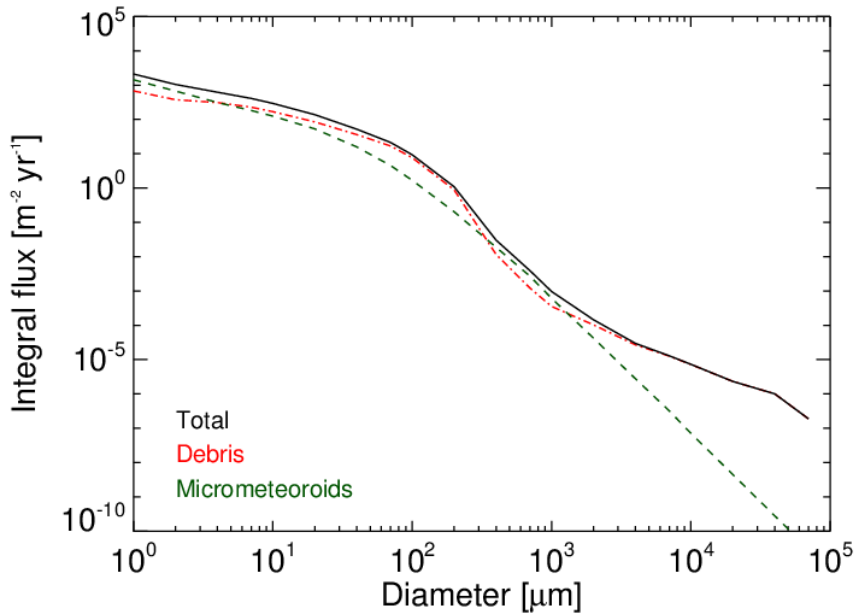

Figure 1: Expected integral flux of micrometeoroids (green dashed curve) and debris (red dash-dotted curve) at the LOFT orbit. For micrometeoroids we assume an average density of 2.5 g/cm$^3$ and velocity of 20 km/s, for debris of 2.8 g/cm$^3$ and 13 km/s.

## 3. RADIATION DAMAGE OF SILICON DRIFT DETECTORS

We summarise in Table 1 the characteristics of the SDDs used for the tests reported in this paper. In the table the comparison is shown with the LAD and WFM SDDs. For all the detectors[8] the thickness is 450 μm and the drift length is 3.5 cm. The SDD is electrically divided into two independent halves. For the detector models with different values of the anode pitch in the two halves, we call "LAD half" the detector half with the largest pitch and "WFM half" the one with the smallest pitch.

Table 1: Characteristics of the SDDs[8] considered in this paper, compared to the ones for the LAD and WFM.

| Model name | Type of measurement | Sensitive area [cm$^2$] | Geometric area [cm$^2$] | Anode pitch [μm] | Number of anodes | Anode volume [cm$^3$] |
|---|---|---|---|---|---|---|
| XDXL1 | Leakage current with 0.8 MeV protons<br><br>CCE with 11.2 MeV protons | 4.35 × 7.02 | 5.52 × 7.25 | 294<br><br>835 | 148<br><br>52 | 4.6 × 10$^{-3}$<br><br>1.4 × 10$^{-2}$ |
| XDXL2 | Leakage current with 11.2 MeV protons<br><br>Hypervelocity particles | 4.35 × 7.02 | 5.52 × 7.25 | 147<br><br>967 | 296<br><br>45 | 2.3 × 10$^{-3}$<br><br>1.5 × 10$^{-2}$ |
| LOFT LAD | | 10.85 × 7.02 | 12.08 × 7.25 | 970 | 112 | 1.5 × 10$^{-2}$ |
| LOFT WFM | | 6.51 × 7.02 | 7.74 × 7.25 | 145 | 448 | 2.3 × 10$^{-3}$ |

### 3.1 Increase in leakage current from the displacement damage

Radiation damage in the bulk of an SDD is caused by the creation of crystal defects with energy states within the semiconductor band-gap. Defects located in the depleted region, with energy levels near the middle of the gap, behave as generation centres that contribute an additional component to the leakage current. The latter can be calculated with the formula[9]

$$\Delta I_{leak} = \alpha \, \Phi_{eq} \, V = \alpha \, k_{eff} \, \Phi \, V \tag{1}$$

where $\alpha$ is the current related damage rate, $k_{eff}$ is the effective hardness factor, $\Phi$ is the fluence at the irradiation energy, and V is the volume of the detector element (collection volume of a single SDD anode).

After the fast recombination of most of the initial interstitial-vacancy (I-V) pairs, created by primary and secondary atom displacements, the remaining I and V point defects migrate and give rise to secondary complex defects by interaction with other crystal defects and impurities. The number of defects and their electrical activity, therefore, changes with time. The net result is a (partial) recover of the leakage current toward its initial value (annealing). This occurs in a way that strongly depends on the device operative temperature since the defect diffusion is thermally activated:

$$\Delta I_{leak,T}(t) = \alpha_T(t) \, \Phi_{eq} \, V \tag{2}$$

Here $\alpha_T(t)$ takes into account the progression of annealing at time t and temperature T. The current related damage rate $\alpha$ in Eq. (1) is defined before annealing starts. In practice, its value can be determined only by measurements at very low temperatures, in such a way that the required experimental time is negligible with respect to the prevailing annealing characteristic time. The value we used for our estimations is $\alpha$ = 11.1×10$^{-17}$ A cm$^{-1}$, measured[9] at -50 °C.

To minimize the leakage current, i.e. to reduce the detector noise, the LOFT SDDs are planned to operate at low temperature, suppressing damage annealing. Due to the constraints at the test facilities, however, we were forced to carry out both irradiations and measurements at room temperature, at which a substantial amount of annealing takes place. We repeatedly measured the SDDs to follow the annealing over a period of several weeks, in such a way to be able to recover the initial damage by applying a suitable model.

For this purpose we selected the annealing model described by Moll et al.[12], which is applicable within the time interval and temperature range of this experimental study. According to this model

$$\alpha_T(t) = \alpha_0 \exp(-t/\tau_I) + \alpha_1 - \alpha_2 \ln(t/t_0) \tag{3}$$

where $\alpha_0$, $\alpha_1$, and $\alpha_2$ are parameters varying slowly with temperature, $t_0$ = 60 s, while

$$\tau_I = [k_{0I} \exp(-E_I/k_B T)]^{-1} \tag{4}$$

has a strong temperature dependence ($k_{0I}$ = 1.2×10$^{13}$ s$^{-1}$, $E_I$ = 1.11 eV, $k_B$ is Boltzmann's constant).

**Irradiation at 11.2 MeV**

We investigated the expected damage from trapped protons by irradiating an XDXL2 SDD at the Proton Irradiation Facility of the accelerator at the Paul Scherrer Institute (PSI) in Villigen (Switzerland). We selected a proton beam with energy spectrum centered at 11.2 MeV and FWHM of ~6 MeV, because this was the available set-up with energy nearest to that used in the SPENVIS estimation (10 MeV). The irradiation was performed in air and at room temperature (~25 °C). The anode pitch is 967 μm, representative of the LAD value (see Table 1). The device under test was not biased during the irradiation. A picture of the set-up at the accelerator is shown in Figure 2.

The average fraction of the flux on the detector surface was 81.4 % of the maximum value, calibrated at steps of 1 cm with a dedicated plastic scintillator before the irradiation. The proton flux was $3.7 \times 10^6$ cm$^{-2}$ s$^{-1}$ and the fluence at the end of the irradiation was $3.1 \times 10^9$ cm$^{-2}$, corresponding to 150 years in orbit at 600 km altitude and 5° inclination (without margin). The uncertainty on the fluence was 2.4 %, due to the time variability of the flux during the irradiation.

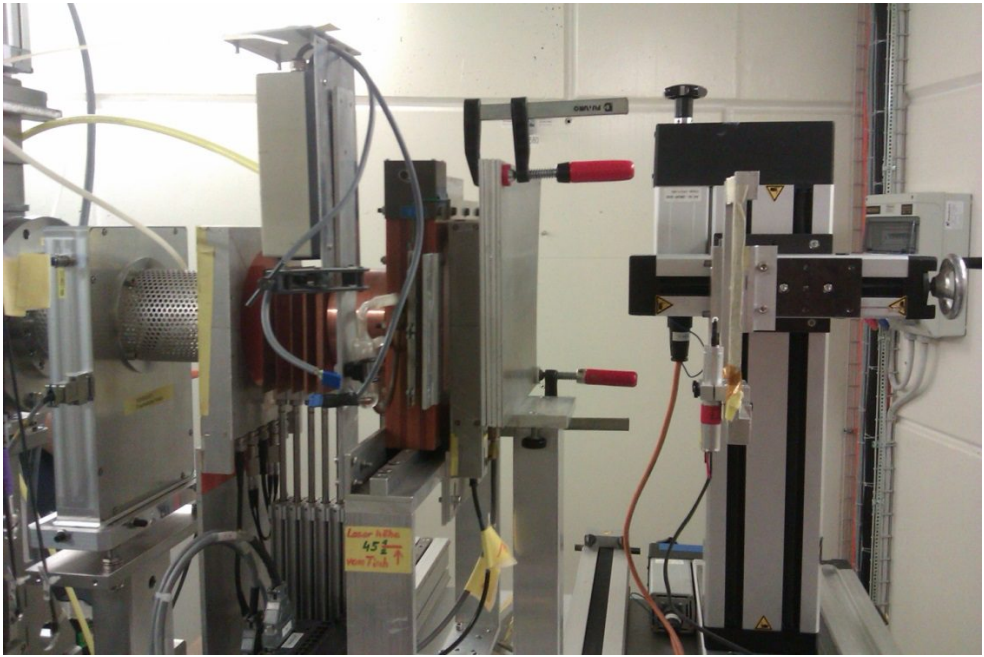

Figure 2: Picture of the experimental set-up at the Proton Irradiation Facility of the PSI during the beam calibration.

**Measurement of the leakage current and annealing after the irradiation (11.2 MeV)**

We measured the leakage current of the anodes in the LAD half, using a dedicated probe station[23], in a time frame between 6.3 days and 70.9 days after the irradiation. The ratio between the measured and expected currents is shown in Figure 3. The first value of the leakage current was measured 6.3 days after the irradiation and was in agreement within 6 % with the estimation from the models of displacement damage[9,10] and annealing[12].

Moreover, we show in Figure 4 that the additional leakage current indeed follows the same variation with temperature as the intrinsic current[13,14,15], as expected. The interested reader may find more details about the experimental set-up and the results in the dedicated publication[15].

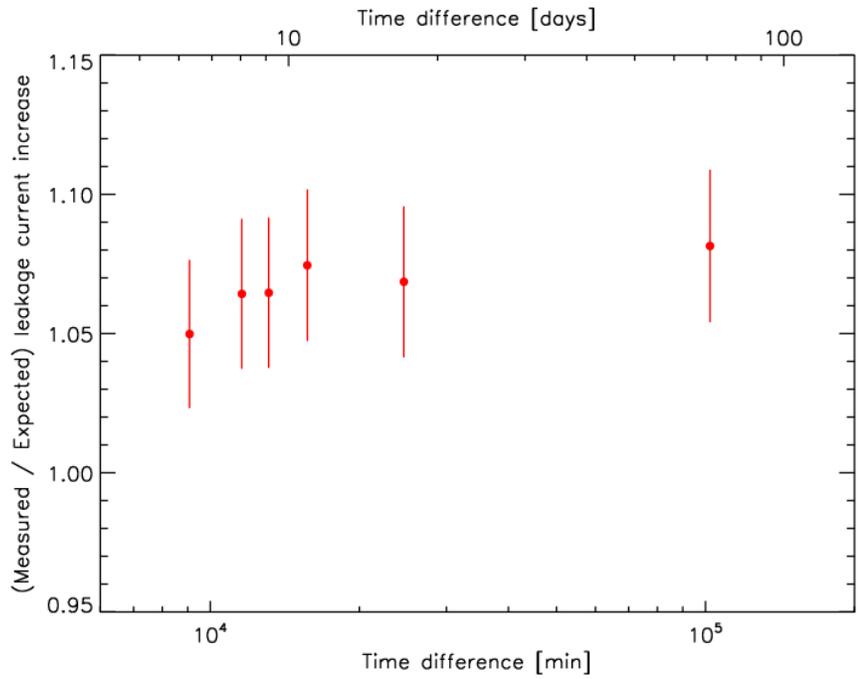

Figure 3: Ratio between the measured and expected increment of leakage current as a function of time after the end of the irradiation.

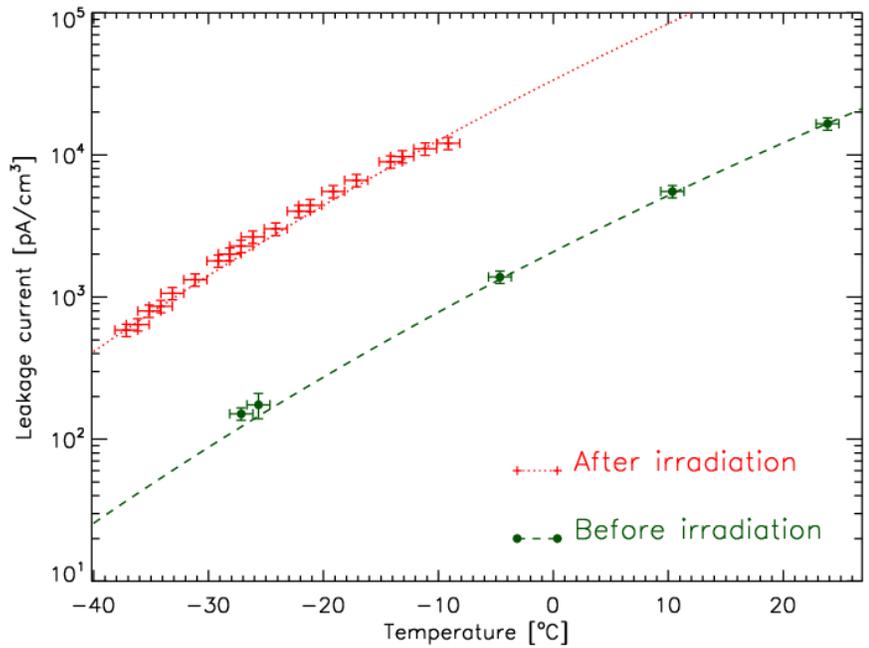

Figure 4: Variation of the leakage current as a function of temperature before (green curve) and after (red curve) the irradiation.

## Irradiation at 0.8 MeV

The radiation damage expected to be produced by soft protons was verified by exposing two SDDs (of XDXL1 and XDXL2 design) to an 838 ± 53 keV proton beam at the accelerator facility[22] of the University of Tübingen (Germany). A picture of one of the tested sensors, placed in the vacuum chamber of the accelerator, is shown in Figure 5. The irradiations were performed in vacuum (residual pressure in the $10^{-6}$ mbar range) at a temperature of ~18 °C with a proton flux of $2.5 \times 10^5$ cm$^{-2}$ s$^{-1}$. The uniformity of the irradiation was within 3% over the whole SDD surface[22]. To be able to easily follow the annealing we planned to expose the detectors to a total fluence of $3.6 \times 10^8$, equivalent to 50 years in orbit for the WFM SDDs (or 75 years for the LAD sensors). The effective fluences in the centre of the SDDs were $3.73 \times 10^8$ cm$^{-2}$ for the XDXL1 prototype and $3.62 \times 10^8$ cm$^{-2}$ for the XDXL2 sensor.

## Measurement of the leakage current and annealing after the irradiation (0.8 MeV)

Soon after being irradiated, the two detectors were stored inside a refrigerated container, at a temperature of -18 °C, to suppress annealing during transportation to Trieste, where the measurement station[23] is located. Extrapolating the Eq. (4) at this temperature one gets $\tau_I \sim 22$ years, which is much larger than the whole duration of the experiment. About one day after irradiation, the SDDs were heated to room temperature and, after 90 minutes, the leakage current was measured for the first time.

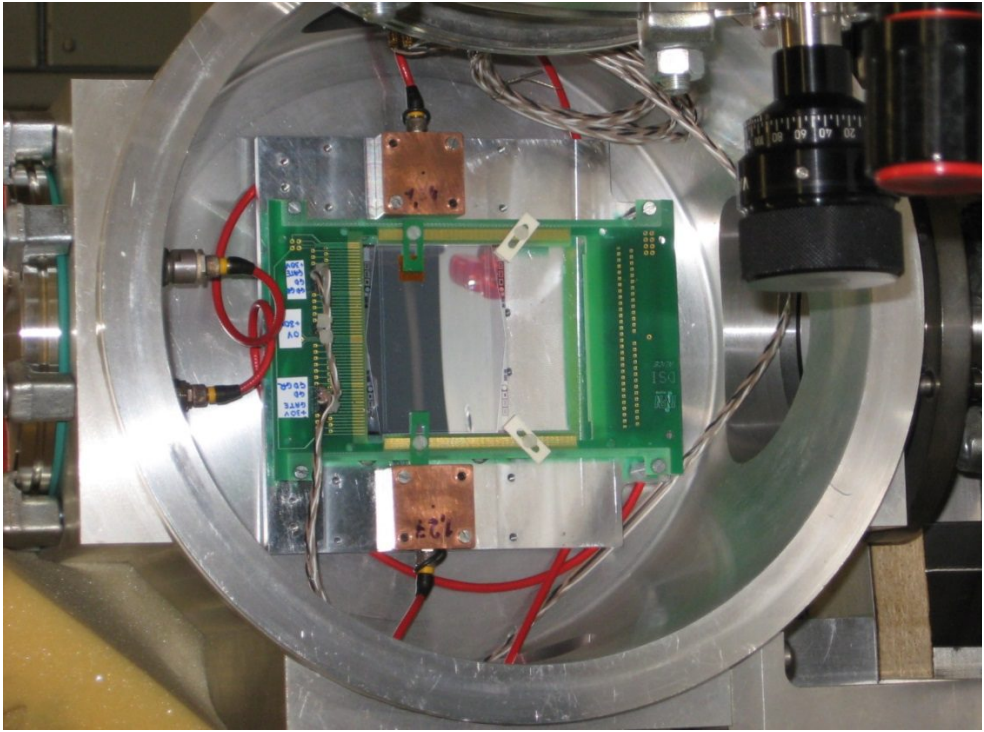

Figure 5: Picture of the SDD in the vacuum chamber for the irradiation with 0.8 MeV protons at the accelerator of the University of Tübingen[22].

The annealing was followed through for about 117 days. Between each measurement the SDDs were not biased and were kept at the laboratory temperature of 24 ± 1 °C. The left panel of Figure 6 reports the time trend of the measured XDXL2 $\Delta I_{leak}$ (solid line with triangular markers) along with the prediction of the annealing model using the parameter values at 21 °C (dashed line).

In the plot the time taken to transport the detectors to Trieste has been neglected due to the small annealing intervened at the storage temperature. The curve showing the leakage current behavior as predicted by the Moll et al.[12] model is plotted starting one day after irradiation because the model is not valid at earlier times. In contrast to what happens at 11.2 MeV, the damage is considerably lower than expected (by a factor greater than 3), as shown in the right panel of

Figure 6. By looking at the trend in the ratio one can also see that the annealing proceeds at a slower rate than predicted: the plot flattens only if a temperature lower than 10 °C is assumed in the annealing model, which was not the case. The same response was found on the XDXL1 detector.

Even if the experimental results differ from the calculations we can safely use our soft proton analysis as a conservative estimation of the in-orbit displacement damage experienced by the LOFT SDDs.

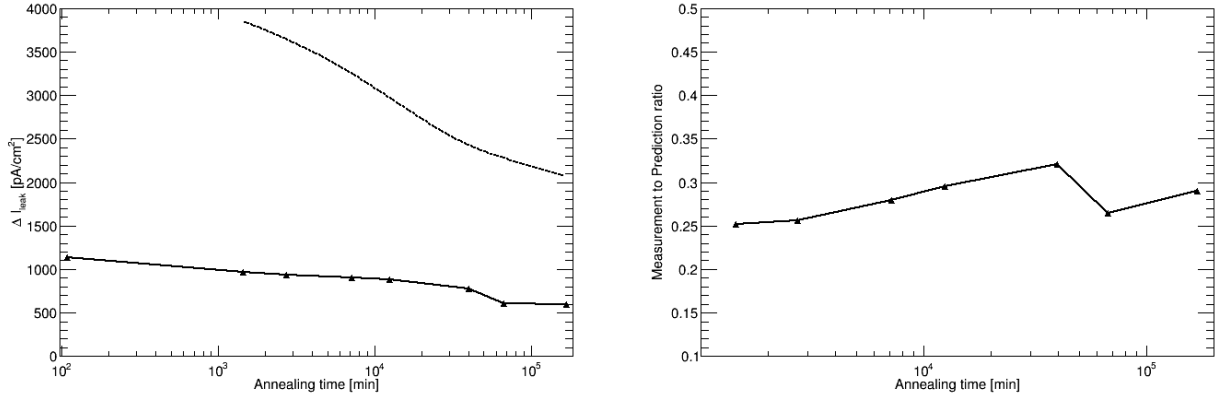

Figure 6: Time evolution of the XDXL2 leakage current. Left: measured values (solid line and triangular markers) and annealing prediction[12] at 21 °C (dashed line). Right: ratio between measurements and prediction.

### 3.2 Variation of the charge collection efficiency

The damage centers produced by the displacement damage in the semiconductor lattice are able to "trap" the charge carriers, thus reducing the charge collection efficiency (CCE) of the device. As reported in the literature[24,25], for small values of the fluence $\Phi$, as is the LOFT case, the reduction of CCE after the irradiation may be approximated at the first order by

$$1 - CCE \sim \beta \, \Phi \, t_c \qquad (5)$$

where $t_c$ is the charge collection time (~5 μs for the SDD under test). The constant $\beta$ in Eq. (5) depends on the specific type of particle used during the irradiation[25] and thus the NIEL hypothesis cannot be applied for the variation of the CCE.

At the LOFT baseline orbit, the expected proton fluence of $3 \times 10^6$ cm$^{-2}$ (including margin) gives an estimated reduction of the CCE of 0.004 % after 4.25 years.

We measured the reduction of the CCE by comparing, before and after the irradiation, the position in the spectrum of the peak of the 5.9 keV manganese $K_\alpha$ line from an $^{55}$Fe source. Since the increased leakage current after the irradiation would worsen the SDD spectral resolution and increase the uncertainty on the reconstruction of the peak position, we accumulated the spectrum at a temperature of about -38 °C. We acquired the spectrum for three positions of the $^{55}$Fe source: near the anodes (at a distance of ~3 mm), at the end of the drift channel (i.e. at ~30 mm from the anodes), and at mid-distance (i.e. at ~15 mm). The source was displaced using a micrometric translation stage. With this method, the CCE is measured basing on the relative distance between the source positions, instead of the absolute locations.

We collimated the source with a slit (1 mm × 10 mm) and a diaphragm (of 400 μm aperture) and we verified that the surface of the beam on the detector was 1.5 mm × 0.75 mm, still on the same anode of the LAD half. The size of the electron cloud, created after the interaction of an X-ray photon, increases during the drift toward the anodes. The expected size of the charge cloud for a drift length of 3.5 cm is ~1 mm[26,27]. Since the anode pitch of the SDD under test was 835 μm, the charge spread on 1 – 3 anodes. In order to ensure a complete collection of the charge, we decided to sum the signals of the triggering anode and of the two adjacent ones, on the left and right. We applied the same method for the three locations of the source along the drift channel.

The device under test was an SDD of the XDXL1 production and only a set of eight contiguous anodes on the LAD half was equipped with a Front-End Electronic[26,27] (FEE), based on discrete components: a SF-51 JFET and an Amptek A250F-NF charge sensitive amplifier. The test equipment (TE) could trigger only on one anode. We subtracted from the signal (sum of the amplitude on the triggering channel and the two adjacent ones) the common mode noise, estimated by averaging the amplitude of the other five channels in the set, and the pedestals of the ADC, independently estimated on each read-out channel with a specific measurement using the same TE.

During the irradiation we provided a proton fluence of $7.9 \times 10^8$ cm$^{-2}$ in a single exposure, with a flux of $1.8 \times 10^6$ cm$^{-2}$ s$^{-1}$. The uncertainty on the fluence, given by the flux variability during the irradiation, was 2.4 %. The PCB was shielded using an aluminum layer ~3 mm thick, able to stop all the protons in the beam, with a rectangular hole in correspondence of the SDD. The average fraction of the beam on the set of eight anodes connected to the FEE was 88.8 % of the maximum. From Eq. (5) we expect a 0.8 % reduction of the CCE for the source position at the end of the drift channel, at a distance of ~30 mm from the anodes.

By repeating the characterisation of the SDD after the irradiation, we measured for the source position at the end of the drift channel a reduction of the CCE of $(0.65 \pm 0.15)$ % , in good agreement with the expected value.

### 3.3 Mitigation strategy for LOFT

Since we verified that the additional leakage current produced by the displacement damage behaves in temperature as the intrinsic leakage current[13,14,15], the mitigation strategy in LOFT is the operation of the SDDs at low temperature. An "active" cooling system, based e.g. on Peltier cells, could not be afforded due to the weight and power consumption. We thus adopted a "passive" temperature control system based on the heat dissipation via dedicated radiators in both the LAD and WFM.

From the prediction of the proton fluence in orbit with the SPENVIS software, we estimate that the LAD operative temperature to fulfil the requirement on the electronic noise ($\sigma_{ENC} \leq 17$ e$^-$ RMS) is lower than -10 °C for the baseline orbit at 550 km altitude and 0° inclination and lower than -52 °C for the more demanding orbit at 600 km and 5°. At these temperatures the damage annealing is negligible.

The expected variation of the CCE is negligible, ~0.004 % at end of life at the baseline orbit, and does not require a mitigation strategy.

## 4. IMPACT OF HYPERVELOCITY PARTICLES ON SILICON DRIFT DETECTORS

While the increase in leakage current produced by the displacement damage on silicon detectors has been extensively studied in literature[9,10,12], few information is available about the effects of the impact of hypervelocity particles such as MMODs. For example, assuming for the debris at the LOFT orbit an average density of 2.8 g/cm$^3$ and a velocity of 13 km/s, the crater produced by a particle of 5 μm diameter has an expected depth of ~16 μm, calculated with the Cour-Palais formula for brittle materials[28]. Similarly, for micrometeoroids the average density and velocity are 2.5 g/cm$^3$ and 20 km/s at the LOFT orbit, thus a particle of 5 μm diameter is expected to produce a crater of ~20 μm depth. Since the total thickness of the passivation and metallisation layers above the silicon bulk is of the order of only 1 μm[29], the craters will extend well into the bulk of the SDD, thus damaging the lattice and increasing the device leakage current.

### 4.1 Experimental set-up

We measured the effect of the impact of hypervelocity particles on the silicon sensors at the Cosmic Dust Accelerator Facility of the Max-Planck Institut für Kernphysik (MPIK) in Heidelberg (Germany). In the campaign we "bombarded" an SDD and five arrays of diodes with iron grains with density of 7.9 g/cm$^3$ and various combinations of diameter and velocity. The SDD had thickness of passive layers and drift length representative of the LOFT detectors (see Table 1). The diodes are simpler structures used to measure the increase of leakage current after each impact that, avoiding the complexity of the SDD layout, allow to reach a clearer understanding of the experiment. A picture of the SDD inside the vacuum chamber of the facility at MPIK is shown in Figure 7. A photo of the diodes is shown in Figure 8.

The leakage current was measured with two Keithley 2657 Source-Measurement Units (SMUs). When "bombarding" the SDD, one SMU was measuring the current on the LAD half and the other on the WFM half. When testing the diodes, one SMU was connected to the anodes and the other to the guard region. The accelerator was operated in "single shot mode", providing the size and velocity of each particle entering in the experimental chamber. With this method we correlated the measured variation of leakage current with the parameters of the particle hitting the detector. During the campaign we shot particles with homogeneous values of the size and velocity on each diode or a region on the SDD. With this choice we produced craters with similar depth on a single diode or on the same area of the SDD.

The size of the dust particles beam was smaller than the SDD surface, thus every particle entering in the experimental chamber hit the SDD. On the other hand, the beam was larger than the active surface of the diodes (light grey in Figure) and a fraction of the particles hit the guard region (whose depleted volume was not well defined and difficult to estimate) or the dead areas of the silicon chips (dark grey in Figure8). In the data analysis we only considered the increase in the diodes leakage current, discarding the impacts on the guard region[29].

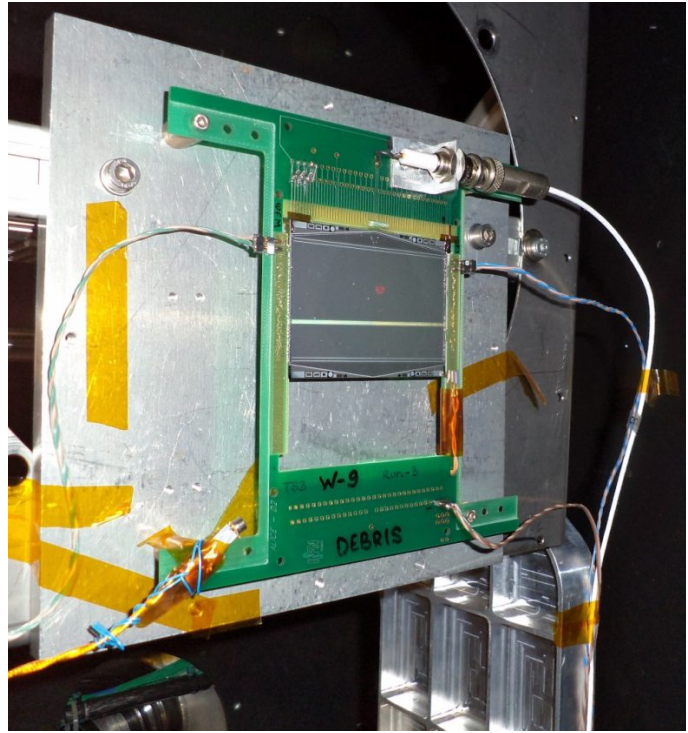

Figure 7: Picture of the SDD under test at the Cosmic Dust Accelerator Facility at MPIK. The red spot on the SDD is due to the laser system used to align the detector with respect to the dust particle beam.

### 4.2 Increase of leakage current after the impacts

An example of the increase of leakage current produced on the SDD (LAD half) by two impacts of particles with a velocity of ~2 km/s and a diameter of ~2 μm is shown in Figure 9. In the plot, the steps in the current value are simultaneous to the particle impacts. In this case the crater depth is ~3 μm, calculated with the Cour-Palais formula[29].

We found in the data analysis that the impact on the SDD of particles with diameter of ~1.1 μm and velocity of ~1.1 km/s did not produce an appreciable increase of the leakage current[29]. The crater depth of these impacts is ~1 μm, which can be considered as a threshold value and is in agreement with the thickness of the passive layers on top of the silicon bulk. After a total number of 46 impacts on the SDD, 24 of which with penetration depth above the threshold estimated above, we measured a leakage current increase of ~43 nA. The detector temperature at the end of the measurement was higher than at the beginning. By correcting for the temperature variation, we found an average increment of ~1.3 nA/impact at 30 °C. The increment is limited to the affected anodes. The interested reader may find more results and details about the campaign with debris in a dedicated publication[29].

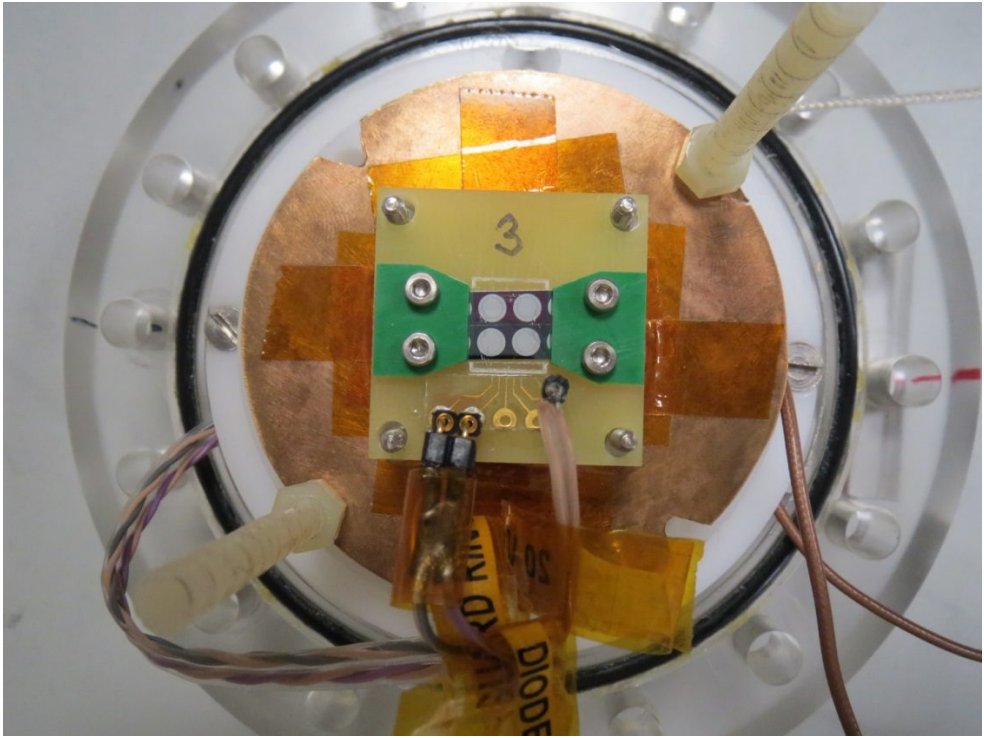

Figure 8: Picture of a set of diodes mounted on the flange of the small experimental chamber at the Cosmic Dust Accelerator Facility.

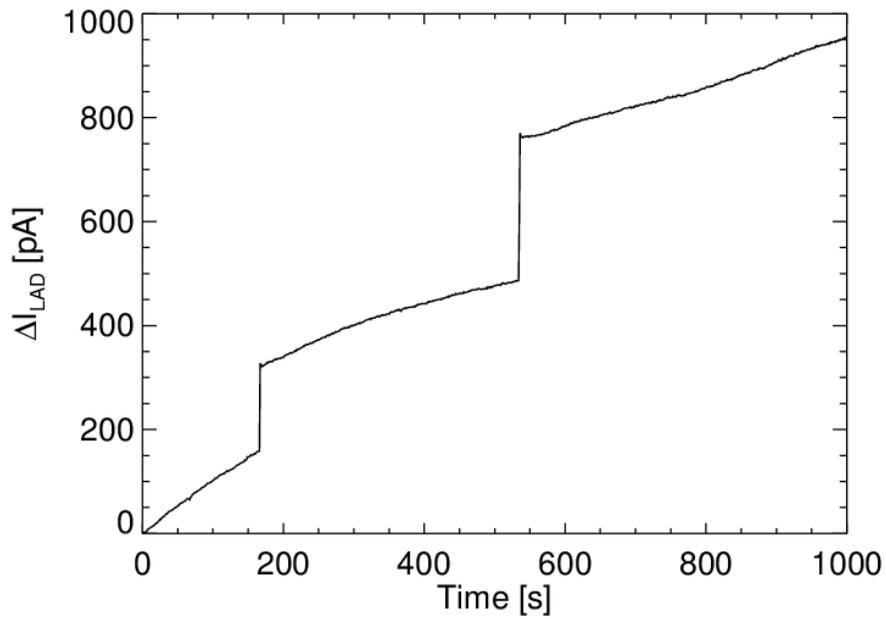

Figure 9: Example of leakage current increase measured on the SDD (LAD half) with two impacts of particles with a velocity of ~2 km/s and a diameter of ~2 μm. The estimated penetration depth is ~3 μm.

### 4.3 Mitigation strategy for LOFT

During the campaign at the MPIK Cosmic Dust Accelerator, we found that the impact of hypervelocity particles on the sensitive surface of the SDD was not destructive, but instead produced a limited increase of leakage current, ~1.3 nA/impact at 30 °C. Although the damage is important and significantly worsens the spectral resolution, it is confined to the impacted and the close neighbor anodes and does not influences the rest of the SDD.

In the LAD design, the SDDs are protected from the MMODs by the lead-glass collimator, whose open channels have a field of view of ~1°. Consequently, the average rate of impacts on the LAD is ~$9.0 \times 10^{-4}$ per SDD per year[29]. Such a low rate does not require any additional mitigation. We estimate that the diameter of the particles able to penetrate through the LAD collimator is ~700 μm, with an expected rate of ~$7 \times 10^{-2}$ particles/yr for the whole LAD[29].

Conversely, on the WFM we estimate ~3.5 impacts/SDD/yr and this could significantly affects the instrument functionality and performance. For this reason we included in the baseline design a 25 μm thick beryllium layer, located directly above the SDDs. In combination with the Multi-Layer Insulator (MLI), composed of 7.6 μm of aluminised Kapton, the beryllium layer acts as a Whipple shield[30], resisting to the impact of micrometeoroids of diameter up to ~20 μm and debris up to ~26 μm, thus reducing the overall expected rate[29] to ~$2.5 \times 10^{-3}$ impacts/SDD/yr. In this analysis we do not take into account the fragmentation of the particles crossing the Whipple Wall.

## 5.  SUMMARY AND CONCLUSIONS

We irradiated a SDD with a proton beam of 11.2 MeV on average at the PIF of the PSI accelerator and we measured an increase of leakage current consistent within 6 % with the model of the displacement damage with the NIEL hypothesis[9,10,11]. The variation of the current in time follows the annealing model reported in literature[12]. After the irradiation, the leakage current produced by the displacement damage has the same variation with temperature[13,14,15] as the intrinsic leakage current of the device, before the irradiation.

We irradiated two SDDs with a proton beam of 0.8 MeV average energy at the accelerator of the University of Tübingen. In this case we found that the measured increase of leakage current is smaller by a factor of ≥3 than expected.

In order to mitigate the increase of leakage current produced by the displacement damage of the LOFT SDDs in orbit, we decided to operate the detectors at a temperature below -10 °C (for the nominal orbit with 550 km altitude and 0° inclination). The operative temperature is reached by means of a "passive" temperature control system, using radiators in both the LAD and WFM.

From the reconstruction of the peak position of the manganese $K_\alpha$ line from an $^{55}$Fe source in three different locations along the drift channel, we measured the variation of the CCE produced by the displacement damage. The variation is in agreement with the model in literature[25]. The expected reduction of CCE in orbit is negligible and does not require a mitigation strategy.

At the Cosmic Dust Accelerator Facility of the MPIK we "bombarded" a SDD and five arrays of diodes with hypervelocity iron particles with various combinations of diameter and velocity. We found that an impact produces a sudden increase of the device leakage current, which is confined to the crater region. We measured on the SDD an average increment of ~1.3 nA/impact at 30 °C. Only particles giving craters with depth above a threshold of ~1 μm are able to produce a measurable increase of leakage current. The threshold crater depth is consistent with the thickness of the passivation and metallisation layers above the silicon bulk.

While in the LOFT design the SDDs of the LAD are protected by the lead glass collimator, with a field of view of ~$3 \times 10^{-4}$ sr, the detectors on the WFM require a dedicated shielding, a 25 μm thick beryllium located just above the detection plane. Together with the MLI above the coded mask, the beryllium layer acts as a Whipple shield[30] and reduces the overall expected rate[29] to ~$2.5 \times 10^{-3}$ impacts/SDD/yr.

## 6. ACKNOWLEDGEMENTS

LOFT is a project funded in Italy by ASI under contract ASI/INAF n. I/021/12/0, in Switzerland through dedicated PRODEX contracts and in Germany by the Bundesministerium für Wirtschaft und Technologie through the Deutsches Zentrum für Luft- und Raumfahrt under grant FKZ 50 OO 1110. The authors acknowledge the Italian INFN CSN5 for funding the Research and Development projects XDXL and REDSOX, and the INFN-FBK collaboration agreement MEMS2 under which the silicon drift detectors were produced. We gratefully acknowledge support of our measurements by the PIF team of PSI lead by W. Hajdas. The LOFT Consortium is grateful to the ESA Study Team for the professional and effective support to the assessment of the mission. This research has made use of NASA's Astrophysics Data System.